# Mg coating induced superconductivity in the FeSe ultrathin film

Wenbin Qiu[2], Zongqing Ma[1, 2*], Yongchang Liu[1], Xiaolin Wang[2], Shi Xue Dou[2]

[1] –Tianjin Key Laboratory of Composite and Functional Materials, School of Materials Science & Engineering, Tianjin University, Tianjin 300072, People's Republic of China

[2] –Institute for Superconducting and Electronic Materials, AIIM, University of Wollongong, Squires Way, North Wollongong, NSW 2500, Australia

**Abstract**

The transition from insulator to superconductor was achieved in the non-superconducting FeSe ultrathin film via simple Mg coating technique in present work. It was found that in non-superconducting FeSe ultrathin film without Mg coating, insulating $\beta$-Fe$_{1-x}$Se phase with iron-vacancy disorders is the main phase and more likely to be the parent phase of FeSe superconducting system. Proper Mg coating on the surface of FeSe films can lead to Mg entering FeSe crystal lattice to fill up these Fe vacancies rather than replace Fe. Simultaneously, additional electron doping is introduced and the electron carrier concentration in this parent phase can be tuned, which is responsible for the SIT and evolution of superconductivity in this system. However, abnormal decrease of electron concentration was found in FeSe film with excessive Mg coating, which brings about the severe degradation in superconducting performance.

* Corresponding author    E-mail: mzq0320@163.com

# 1. Introduction

Compared with bulk crystals, FeSe in the form of thin films has generated wide research interest as large enhancement in superconducting performance is obtained. In particular, the monolayer (ML) FeSe thin film on $SrTiO_3$ substrate exhibits unexpected high-$T_c$ superconductivity close to boiling temperature of liquid nitrogen (77K)[1-5]. Unfortunately, there has been no report of multilayer FeSe films showing a signature of superconductivity. In fact, superconductor-insulator transition (SIT) occurs when the thickness of FeSe thin film increases from monolayer to two or more layers [1, 3]. On the other hand, in the FeSe thin films (in the bulk nature) on various substrates such MgO, $CaF_2$ and $SrTiO_3$ [6-10], SIT and evolution of superconductivity were also observed depending on the thickness. It was reported that suppression on superconductivity occurs in these films as thickness decreases and there is a critical thickness for SIT in these films [7, 8, 10]. Various parameters might induce SIT, and clearly transitions were found by tuning disorder [6, 7] or magnetic field [8]. Till now the underlying mechanism of SIT is still under controversy. Understanding the nature of SIT in FeSe film can provide a valuable hint to reveal the intrinsic mechanism of superconductivity in this system.

According to our recent research [10], the concentration of electron carrier, corresponding to the Fe-vacancy disorder in FeSe thin films with different thickness is the intrinsic factor determining the SIT and the evolution of the superconductivity, rather than other thickness-related factors, such as internal pressure effect [11], interface effect [12, 13]. The amount of Fe-vacancy disorders was found to vary in the

FeSe films with different thickness, which obviously changes the concentration of electron carrier. As a result, electron doping was introduced, determining SIT and the evolution of superconductivity in FeSe films. Very recently, it was also found that superconductivity emerges in multilayer FeSe films when enough electrons doping was introduced via potassium coating [14]. Both results suggest that electrons doping could be the crucial factor for SIT in FeSe superconducting system.

In this point of view, it should be strong evidence to the speculation mentioned above if the transition from insulator to superconductor can be achieved in non-superconducting FeSe thin films with thickness below critical value by introducing external electron doping. Moreover, in contrast to the instability of FeSe multilayer in atmosphere, FeSe thin film can be stable in atmosphere and thus could act as a satisfying media for various characteristics or physical measurements to explore SIT phenomenon as well as understand the nature of electron doping in this unique FeSe superconducting system.

In present work, the transition from insulator to superconductor was achieved in the FeSe ultrathin film via Mg coating technique. Proper Mg coating on the surface of FeSe films can introduce electron doping and induce the variation of electron carrier concentration, which is responsible for the SIT and evolution of superconductivity in this system. This is the first time to achieve transition from insulator to superconductor in non-superconducting FeSe ultrathin films with thickness below critical value via a simple metal coating process.

## 2. Experimental

FeSe films are grown on $CaF_2$ (100) substrate (lattice parameter $a$ = 5.462 Å) by PLD technique (Nd: YAG, 355 nm, 10 Hz, 2 W output). Substrate temperature is chosen as 300 ˚C and the entire experiment is carried out in a close chamber with vacuum better than $5 \times 10^{-4}$ Pa. Home-made FeSe pellet which is prepared from commercial available Fe and Se powder serves as target for FeSe sputtering. In order to check the influence of Mg content on FeSe films, the thickness of FeSe films is strictly fixed at 40 nm which shows non-superconducting. Right after the deposition of FeSe film, Mg coating is introduced by directly switching to Mg target holder. By controlling the deposition time of Mg as 0min, 3 min, 10 min and 20 min, a batch of Mg-coated FeSe films are obtained and denoted as #FM0, #FM1, #FM2 and #FM3, respectively. Both of the electrical resistivity and Hall coefficient measurements are carried out in a 9 T physical property measurement system (PPMS). XRD *θ-2θ* scan is used to characterize the crystal structure of films. For checking the surface morphology and chemical composition, scanning electron microscopy (SEM, JEOL JSM-6490LV) and high-accuracy energy dispersive X-ray spectrum (EDX) detector (Oxford Instruments X-MaxN 80) is employed.

## 3. Results and discussion

Fig. 1(a) shows the temperature dependence of electrical resistivity for all films (#FM0, #FM1, #FM2 and #FM3) in magnified temperature range, while full range up to 300 K is plotted in the inset of Fig. 1(a). As for uncoated FeSe film with thickness about 40 nm (#FM0), insulator-like behavior is observed in R-T curve especially below 50K, whereas for Mg coated films (#FM1, #FM2 and #FM3), their R-T curves behave in a metallic way. One can see that clear insulator-superconductor transition is obtained from uncoated FeSe film to FeSe film (#FM1) even with only small amount of Mg coating. Keep increasing the amount of Mg coating, $T_c^{onset}$ increases from 6.3 K for #FM1 to 9.6 K for #FM2. This is the first time to achieve transition from insulator to superconductor in FeSe ultrathin films with thickness below critical value via a simple metal coating process. Moreover, these Mg coated films are quite stable in atmosphere and can be employed to various characteristics and properties measurement to further study the nature of the underlying mechanism of SIT. Interestingly, with the amount of Mg coating further increasing, degradation in $T_c^{onset}$ is observed from 9.6 K for #FM2 to 7.3K for #FM3.

In order to further investigate the upper critical magnetic field ($H_{c2}$) of our Mg coated films with better superconductivity (#FM2 and #FM3), measurements of R-T are executed by applying different magnetic fields, and such an example is shown in Fig 1(b). The transition shifts to lower temperatures under higher field and $\Delta T_c$ gets wider, which is a common characteristic in type-II superconductors. $H_{c2}$ is determined by linear extrapolation fitting of $T_c^{mid}$, as shown in Fig 1(c). The estimated $H_{c2}$ in these two superconducting FeSe films are about 30.4 T and 21.4 T for #FM2 and #FM3, respectively, which is comparable to the best value of $H_{c2}$ in the uncoated superconducting FeSe films with thickness above critical value, as reported in previous study [10, 11].

As shown above, insulator-superconductor transition in FeSe thin films can be achieved by Mg coating, thus it is of great importance to understand its intrinsic mechanism. It is well known that charge carrier concentration is one of the most crucial parameters that influence the superconductivity in iron-based superconductors [15]. As Mg belongs to the group of alkaline earth, abundant electron carriers are supposed to be provided through Mg coating on the surface of FeSe films. Hereby hall measurements are performed to reveal the state of charge carrier in all samples. In Fig. 2, temperature dependences of Hall coefficient ($R_H$) are illustrated. $R_H$ is defined as $R_H = \rho_{xy}/B$, where $\rho_{xy}$ stands for Hall transverse resistivity and B is designated field under fixed temperatures ranging from 20 K to 300 K. In the case of sample #FM0, some fluctuation is observed in $R_H$ value which is probably due to the poor response of transport $R_{xy}$ signal detected in insulator-like FeSe film. As temperature goes down from room temperature to 50K, $R_H$ slightly increases at minus side and exhibits a broad peak at about 50 K. Then the absolute value of $R_H$ steeply decrease and even reaches the value above 0 as temperature further goes down from 50K to 20K. $R_H$ value changing from negative to positive at low temperature means that the dominant charge carriers change from electrons to holes in this non-superconducting FeSe film. The electron concentration hits a lower bound near 50 K and the inflection point in Hall measurement is in good agreement with R-T dependence where the resistivity of sample #FM0 shows a significant upturn after 50 K. Hence, we predict that insulator behavior in our 40 nm thick non-superconducting FeSe film is induced by the transformation in electronic state at 50 K, and superconductivity is supposed to emerge once this transformation is suppressed by introducing more electron charge carrier. Actually, in Chen et al's work [16], they proposed that an insulating β-$Fe_{1-x}$Se phase with particular iron-vacancy disorder in Fe-Se system could be the parent phase of FeSe superconducting system instead of β-$Fe_{1+\delta}$Te phase. Our results here are quite consistent with their prediction. In our very recent work [10], it was found that β-$Fe_{1-x}$Se phase with iron-vacancy disorders is the main phase in our non-superconducting FeSe film #FM0 based on XRD and EDX results, which brings about the insulator-like behavior in R-T curve and abnormal change in the $R_H$-T curve (see Fig. 1a and Fig. 2), and can serve as parent phase of FeSe superconducting system. On the other hand, $R_H$ in all the Mg coated FeSe films have negative value at the temperature above $T_c$, indicating that electrons are dominate carrier type in all of our Mg-coated FeSe films. From room temperature to 150 K, the $R_H$ of all the samples is almost temperature independent. Below that, the absolute value of $R_H$ starts to increase, implying the multiband nature in typical FeSe system. As the carrier concentration $n$ is inversely proportional to $R_H$, the increasing $R_H$ in minus side at lower temperature region represents that the electron carrier density diminishes with decreasing temperature. Comparing with the values of $R_H$ in these Mg coated FeSe films, one can see that more electrons are introduced and the corresponding electron carrier concentration increases as more Mg coating was deposited on the surface of FeSe films, except that the #FM3 film with the most amount of Mg coating possesses unexpected minimum electron carrier concentration. All these results are corresponding to the decaying behavior in the temperature dependence of resistivity as well as evolution of $T_c^{onset}$ as shown in Fig. 1(a), indicating that variation of carrier concentration via Mg coating is the crucial factor determining the transition from insulator to superconductor and evolution of superconductivity in FeSe films.

Typical XRD patterns of four FeSe films are displayed in Fig 3(a) ranging from 10 to 80 degree. One can see that highly *(00l)* oriented FeSe texture based on PbO structure exists in all 5 samples. Noteworthily, no new phase is detected after Mg coating. It strongly manifests that instead of reacting with FeSe, Mg enters into the lattice structure of FeSe. This speculation is diverse from the result raised in Mg-doped FeSe bulks [17] where MgSe new phase is formed with Mg addition, representing absolutely different doping mechanism of Mg in FeSe films and bulks. In Mg-doped FeSe bulks made by solid-state-sintering, the particles of Mg addition is very small and reactive, which can contact FeSe particles well and easily react with FeSe forming MgSe secondary phase at relatively higher sintering temperature. Oppositely, in present work, Mg particles only homogeneously deposit on the surface of FeSe films fabricated by PLD technique at relatively lower deposition temperature (300 ˚C), so that Mg can readily enter into FeSe lattice, and bring about novel properties in this system.

Fig. 3(b) shows the enlarged interval near β-FeSe (001) peak for prepared films. No obvious shift for diffraction peak position is recognized in all Mg coated FeSe films compared with uncoated one, implying that there is no change in the lattice parameter *c* of β-FeSe after Mg coating even though Mg is confirmed to enter its crystal lattice as discussed above. Since the uncoated FeSe film prepared here contains abundant Fe vacancies (the Fe/Se ratio is about 1.00:1.09 in this film), the Mg that enters the FeSe crystal lattice mainly fills up these vacancies instead of replacing Fe. Consequently, no obvious lattice distortion can be introduced and the corresponding lattice parameter almost keeps unchanged after Mg coating. Simultaneously with this Mg filling up Fe vacancies, additional electron carrier can be introduced and higher superconductivity will be obtained, as shown in Fig. 1a and Fig. 2. However, why more Mg coating in film #FM3 leads to abnormally low electron carrier concentration and then the depression of superconductivity is still unclear and needs to be further investigated.

## 4. Conclusion

The transition from insulator to superconductor was achieved in the FeSe ultrathin film via simple Mg coating technique. It was found that in non-superconducting FeSe ultrathin film without Mg coating, insulating $\beta$-$Fe_{1-x}$Se phase with iron-vacancy disorders is the main phase and more likely to be the parent phase of FeSe superconducting system. Proper Mg coating on the surface of FeSe films can lead to Mg entering into the FeSe lattice structure instead of reacting with FeSe to form new phase. The Mg that enters the FeSe crystal lattice mainly fills up Fe vacancies rather than replace Fe. Simultaneously, additional electron carrier can be introduced and higher superconductivity will be obtained. However, abnormal decrease of electron concentration was found in FeSe film with excessive Mg coating, which brings about the severe degradation in superconducting performance.

**Acknowledgement** *This work is supported by the Australian Research Council (Grant No. DE140101333). The authors are also grateful to the China National*

*Funds for Distinguished Young Scientists (Grant No. 51325401) and National Natural Science Foundation of China (Grant No. 51302186 and 51574178). The authors acknowledge use of facilities within the UOW Electron Microscopy Centre.*

Figure 1

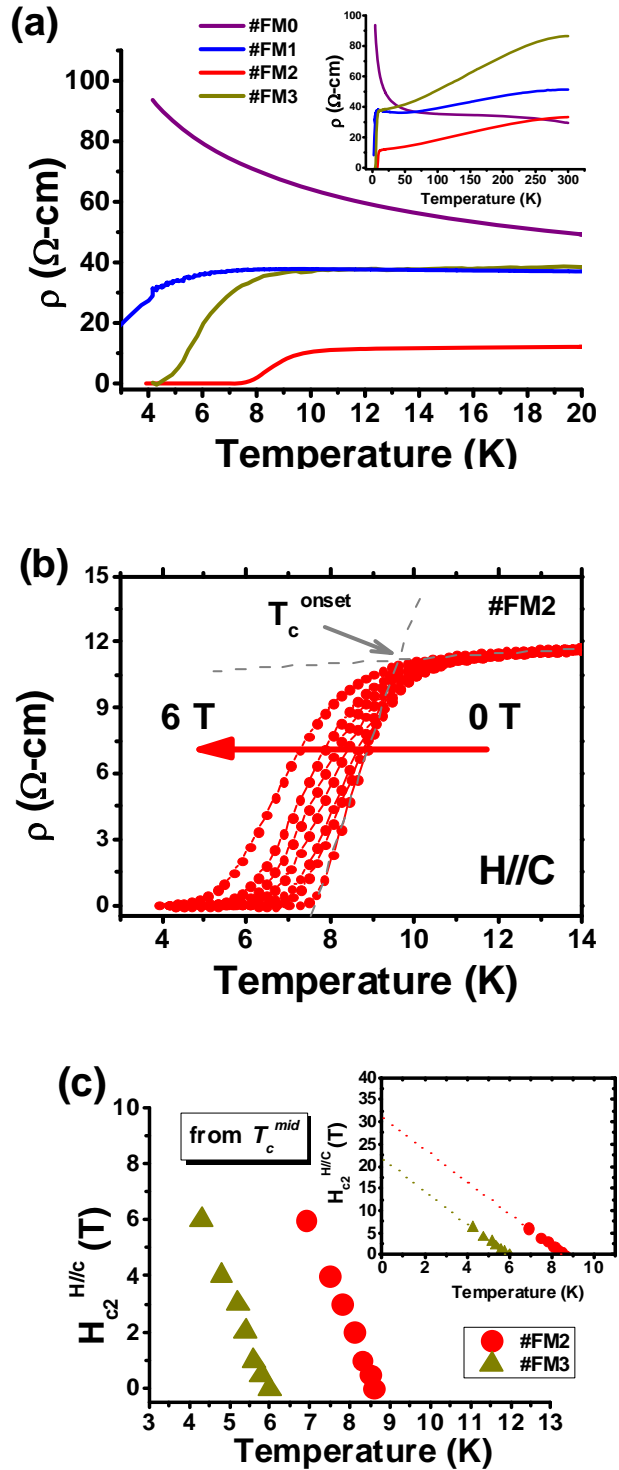

FIG. 1 (a) Temperature dependences of resistivity (R-T curves) for Mg-coated FeSe films with different deposition time of Mg (0 min, 3 min, 10 min and 20 min, corresponding to #FM0, #FM1, #FM2 and #FM3, respectively) ranging from 4 K to 20 K. Inset is the full range R-T data from room temperature to 4.2 K. (b) R-T curves for Mg coated FeSe film #FM2 under different magnetic fields up to 6 T. (c) Plot of upper critical field ($H_{c2}$) as a function of $T_c^{mid}$ for two Mg coated FeSe films with better superconductivity. The linear extrapolations to T = 0 K are shown in the inset.

Figure 2

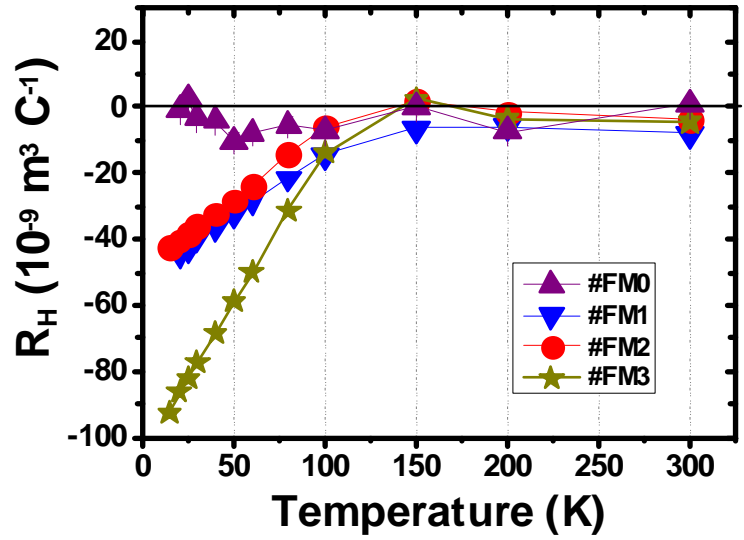

FIG 2. Hall coefficient $R_H$ as a function of temperature of Mg coated FeSe films. $R_H$ is determined as $R_H = \rho_{xy}/B$, dedicating the slope of hall transverse resistivity $\rho_{xy}$ at designated field B.

Figure 3

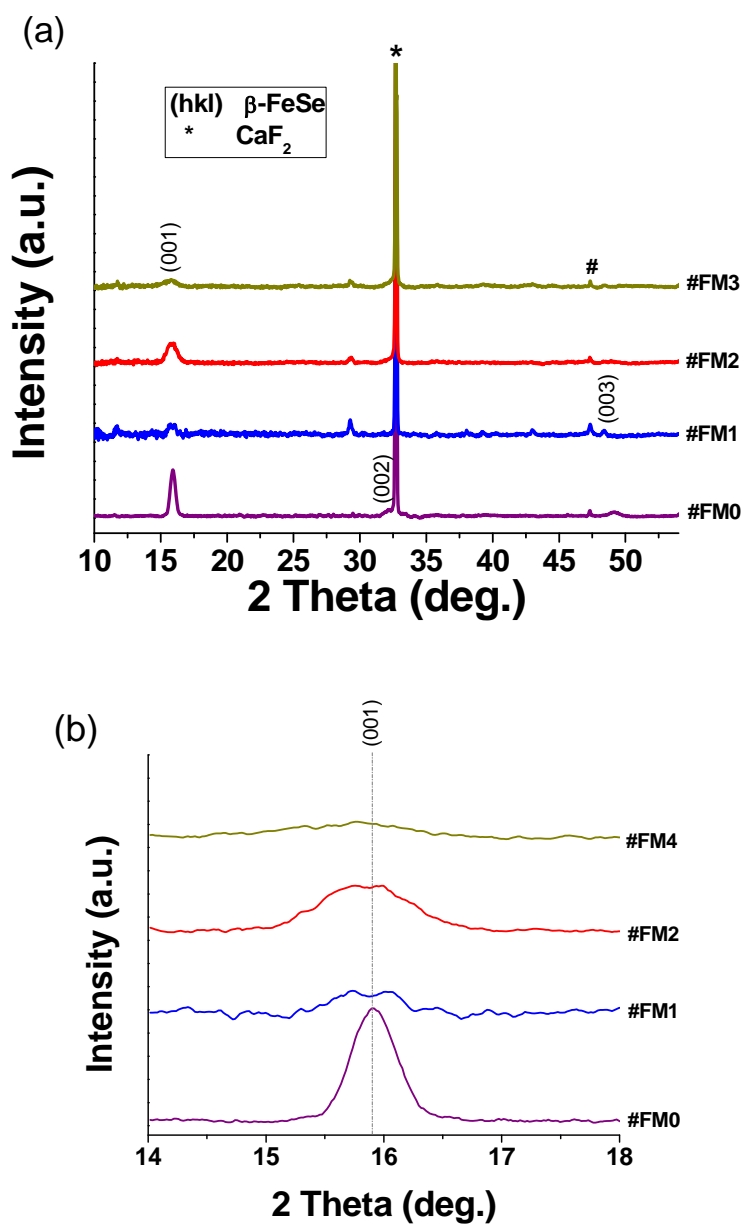

FIG 3. (a) XRD patterns of Mg coated FeSe films grown on CaF$_2$ (100) substrate. Number signs (hkl) represent β-FeSe phases and the pound sign stands for unidentified peak. (b) Magnified interval near β-FeSe (001) peak and the locations of peaks are marked by dashed lines.